\documentclass[12pt]{article}

\usepackage{graphicx}
\input{epsf}

\usepackage{amssymb}
\usepackage{amsmath}

\begin{document}

\title{RS model with the small curvature and Bhabha scattering at the ILC}

\author{ A.V. Kisselev \thanks{E-mail: alexandre.kisselev@ihep.ru}\\
{\small Institute for High Energy Physics, Protvino, Russia}}

\date{}

\maketitle

\begin{abstract}
The Randall-Sundrum (RS) model with the small curvature is
studied. In such a scheme the mass spectrum of Kaluza-Klein (KK)
gravitons is similar to that in a model with one extra flat
dimension. The gravity effects in the Bhabha scattering  at the
energy 1 TeV are estimated. The calculations are based on the
analytical formula which describes virtual graviton contributions.
It takes into account both a discrete character of the mass
spectrum and nonzero widths of the KK gravitons.
\end{abstract}


\section{RS model with the small curvature}
\label{sec:RS_model}

The Randall-Sundrum (RS) model~\cite{Randall:99} is realized in a
slice of the AdS$_5$ space-time with the background warped metric:
\begin{equation}\label{metric}
ds^2 = e^{2 \kappa (\pi r - |y|)} \, \eta_{\mu \nu} \, dx^{\mu} \,
dx^{\nu} + dy^2 \;,
\end{equation}
where $y = r \theta$ ($-\pi \leqslant \theta \leqslant \pi$), $r$
being the ``radius'' of extra dimension, and  $\eta_{\mu \nu}$ is
the Minkowski metric. The points $(x_{\mu},y)$ and $(x_{\mu},-y)$
are identified.

The parameter $\kappa$ defines a 5-dimensional scalar curvature of
the AdS$_5$ space. Namely, the Ricci curvature invariant
$\mathcal{R}^{(5)} = -20 \, \kappa^2$. That is why, in what
follows we will call $\kappa$ ``curvature''.

There are two 3D branes in the model with equal and opposite
tensions located at the point $y = \pi r$ (called the \emph{TeV
brane}) and point $y = 0$ (referred to as the \emph{Plank brane}).
If $k > 0$, the tension on the TeV brane is negative, whereas the
tension on the Planck brane is positive. All the SM fields are
confined to the TeV brane, while the gravity propagates in all
five dimensions.

It is necessary to note that the metric \eqref{metric} is chosen in
such a way that 4-dimensional coordinates $x_{\mu}$ are Galilean on
the TeV brane where all the SM field live, since the warp factor is
equal to unity at $y = \pi r$.%
\footnote{To get a right interpretation, one has to calculate the
masses on each brane in the Galilean coordinates with the metric
$g_{\mu \nu} = (-1,1,1,1)$ (for details, see
Ref.~\cite{Boos:02}).}

By integrating a 5-dimensional action over variable $y$, one gets an
effective 4-dimensional action, that results in the ``hierarchy
relation'' between the reduced Planck scale $\bar{M}_{\mathrm{Pl}}$
and 5-dimensional gravity scale $\bar{M}_5$:
\begin{equation}\label{RS_hierarchy_relation}
\bar{M}_{\mathrm{Pl}}^2 = \frac{\bar{M}_5^3}{\kappa} \left( e^{2 \pi
\kappa r} - 1 \right) \;.
\end{equation}
The reduced 5-dimensional Planck mass $\bar{M}_5$ is related to
the Planck mass by $M_5 = (2\pi)^{1/3} \bar{M}_5 \simeq 1.84
\bar{M}_5$.

From the point of view of a 4-dimensional observer located on the
TeV brane, there exists an infinite number of graviton KK
excitations with masses
\begin{equation}\label{graviton_masses}
m_n = x_n \, \kappa, \qquad n=1,2 \ldots \;,
\end{equation}
where $x_n$ are zeros of the Bessel function $J_1(x)$.

The interaction Lagrangian on the visible brane looks like the
following:
\begin{equation}\label{Lagrangian}
\mathcal{L} = - \frac{1}{\bar{M}_{\mathrm{Pl}}} \, T^{\mu \nu} \,
G^{(0)}_{\mu \nu} - \frac{1}{\Lambda_{\pi}} \, T^{\mu \nu} \,
\sum_{n=1}^{\infty} G^{(n)}_{\mu \nu} +
\frac{1}{\sqrt{3}\Lambda_{\pi}} \, T^{\mu}_ {\mu} \, \phi \;.
\end{equation}
Here $T^{\mu \nu}$ is the energy-momentum tensor of the matter on
this brane, $G^{(n)}_{\mu \nu}$ is a graviton field with the
KK-number $n$, and $\phi$ is a scalar field called radion. The
parameter $\Lambda_{\pi}$ in Eq.~\eqref{Lagrangian},
\begin{equation}\label{lambda}
\Lambda_{\pi} = \bar{M}_5 \,\left( \frac{\bar{M}_5}{\kappa}
\right)^{\! 1/2} ,
\end{equation}
is a physical scale on the TeV brane.

In most of the papers which treat the RS model,%
\footnote{Including the original one \cite{Randall:99}.}
the hierarchy relation is of the form
\begin{equation}\label{hierarchy_relation_old}
\bar{M}_{\mathrm{Pl}}^2 = \frac{\bar{M}_5^3}{\kappa} \left( 1 -
e^{-2 \pi \kappa r} \right) \;,
\end{equation}
while the curvature and 5-dimensional gravity scale are chosen to
be of the order of the Planck mass:
\begin{equation}\label{scale_relation_uncorr}
\kappa \sim \bar{M}_5 \sim \bar{M}_{\mathrm{Pl}} \;.
\end{equation}
In such a scheme, one obtains a series of massive graviton
resonances in the TeV region which interact rather strongly with
the SM fields, since $\Lambda_{\pi} \sim 1$ TeV on the TeV brane.
The lightest modes of the KK graviton have masses around 1 TeV.
Thus, an experimental signature of the ``\emph{large curvature
option}'' of the RS model is the real or virtual production of the
massive KK graviton resonances.

There are lower bounds on the mass of the lightest KK mode, coming
from the measurements of $e^+e^-$ and $\gamma \gamma$ final states
on the Tevatron~\cite{Tevatron_limits}:
\begin{eqnarray}\label{Tevatron_limits}
m_1 &>& 875 \mathrm{\ MeV} \quad (\mathrm{CDF}) \;,
\nonumber\\
m_1 &>& 865 \mathrm{\ MeV}  \quad  \; (\mathrm{D0}) \;.
\end{eqnarray}
The LHC search limit on $m_1$ is \cite{LHC_limits}
\begin{equation}\label{LHC_dijets}
m_1 \sim 0.7- 0.8 \mathrm{\ TeV}
\end{equation}
for the di-jet mode and integrated luminosity $\mathcal{L}= 0.1
\mathrm{\ fb}^{-1}$. For the di-photon mode and $\mathcal{L}= 10
\mathrm{\ fb}^{-1}$, the limit looks like~\cite{LHC_limits}
\begin{equation}\label{LHC_diphoton}
m_1 \sim 1.31 - 3.47 \mathrm{\ TeV} \;.
\end{equation}

In the case of large curvature \eqref{scale_relation_uncorr}, the
size of the $AdS_5$ slice is extremely small:
\begin{equation}\label{radius}
r \simeq 60 \, l_{\mathrm{Pl}} \;,
\end{equation}
where $l_{\mathrm{Pl}}$ is a Planck length. Thus, in order to
explain the huge value of $\bar{M}_{\mathrm{Pl}}$, one has to
introduce large mass scales $\bar{M}_5$, $\kappa$, as well as
$1/r$. In other words, \emph{hierarchy problem is not solved, but
reformulated} in terms of the new parameter related to the size
of the bulk along the extra dimension.%
\footnote{A similar shortcoming exists in models with large extra
dimensions.}

In the present paper we will consider the ``\emph{small curvature
option}'' of the RS model~\cite{Kisselev:05,Kisselev:06}. In what
follows, the 5-dimensional reduced Planck mass $\bar{M}_5$ is
taken to be one or few TeV. Following
Ref.~\cite{Kisselev:05}-\cite{Giudice:04}, we chose the parameter
$\kappa$ to be $900 \mathrm{\ MeV} - 1\mathrm{\ GeV}$. These
values of $\kappa$ obey the bounds derived in
Ref.~\cite{Kisselev:05}:
\begin{equation}\label{curvature_limits}
10^{-5} \leqslant \frac{\kappa}{\bar{M}_5} \leqslant 0.1 \;.
\end{equation}
Then the mass of the lightest KK excitation $m_1 \simeq 3.4 -
3.8$~GeV.

On the contrary, the value of the mass scale
$\Lambda_{\pi}$~\eqref{lambda} is rather large:
\begin{equation}\label{lambda_enum}
\Lambda_{\pi} = 100 \left( \frac{M_5}{\mathrm{TeV}} \right)^{3/2}
\left( \frac{\mathrm{100 \ MeV}}{\kappa} \right)^{1/2} \!
\mathrm{TeV} \;.
\end{equation}
It immediately follows from Eqs.~\eqref{Lagrangian},
\eqref{lambda_enum} that there is no problem with the radion field
$\phi$ in our scheme, since its coupling to the SM fields ($\sim
1/\sqrt{3}\Lambda_{\pi}$) is strongly suppressed.

As for the massive gravitons, their couplings are also defined by
the scale $\Lambda_{\pi}$~\eqref{lambda_enum}. However, the
smallness of the coupling is compensated by the large number of
the gravitons that can be produced in any inclusive process. As a
result, magnitudes of cross sections will be defined by the
5-dimensional gravity scale $\bar{M}_5$ and collision energy
$\sqrt{s}$ (see below).

Thus, we have an infinite number of low-mass KK resonances with
the small mass splitting, in contrast with the standard RS
scenario~\eqref{scale_relation_uncorr}.%
\footnote{Bounds \eqref{Tevatron_limits}-\eqref{LHC_diphoton}
\emph{are not valid} for the small curvature scenario of the RS
model.}
Nevertheless, due to the warp geometry of the $AdS_5$ space-time,
the RS model with the small curvature differs significantly from
the ADD model~\cite{Arkani-Hamed:98} (at least, at $\kappa \gg
10^{-22}$ eV), as was demonstrated in Ref.~\cite{Kisselev:06}.

Let us stress that the hierarchy relation
\eqref{RS_hierarchy_relation} explains the value of the
4-dimensional Planck mass without introducing large mass scales,
if one put $r = 9.7/\kappa \simeq 10 \mathrm{\ GeV}^{-1}$.

Recently, the small curvature option of the RS model has been
checked by the DELPHI Collaboration. The gravity effects were
searched for by studying photon energy spectrum in the process
$e^+e^- \rightarrow \gamma + E_{\perp}\hspace{-6mm} \diagup
\hspace{2mm}$. No deviations from the SM prediction were seen. As
a result, the following bound has been obtained~\cite{LEP_limit}:
\begin{equation}\label{n=1_limit}
M_5 > 1.69 \mathrm{\ TeV} \pm 3 \% \;,
\end{equation}
that corresponds to the reduced 5-dimensional scale $\bar{M}_5 >
0.92$ TeV (see the relation between $M_5$ and $\bar{M}_5$ after
Eq.~\eqref{RS_hierarchy_relation}).

Note that this limit could not be inferred from the limits already
given for larger than two flat dimensions owing to the totally
different spectrum of the photon~\cite{LEP_limit}.

\section{Virtual KK gravitons in the Bhabha scattering}
\label{sec:virtual_gravitons}

Following Ref.~\cite{Kisselev:07}, we will consider now the Bhabha
process at the ILC collider~\cite{ILC_projects} mediated by
massive graviton
exchanges:%
\footnote {The processes $e^+e^- \rightarrow \mu^+ \mu^-$, $e^+e^-
\rightarrow \gamma \gamma$ are also promising reactions.}
\begin{equation}\label{process}
e^+ \, e^- \rightarrow G^{(n)} \rightarrow e^+ \, e^-  \;.
\end{equation}

The collision energy $\sqrt{s}$ is taken to be 1 TeV. It means
that we are working in the following region:
\begin{equation}\label{energy_region}
\Lambda_{\pi} \gg \sqrt{s} \sim M_5 \gg \kappa \;.
\end{equation}

The matrix element of the process~\eqref{process} looks like
\begin{equation}\label{matrix_element}
\mathcal{M} = \mathcal{A} \, \mathcal{S} \;.
\end{equation}
The fist factor in Eq.~\eqref{matrix_element} is the contraction
of the the tensor part of the graviton propagator, $P^{\mu \nu
\alpha \beta}$, and the energy-momentum tensor of the electron,
$T_{\mu \nu}^{e}$:
\begin{equation}\label{tensor_contraction}
\mathcal{A} = T_{\mu \nu}^e \, P^{\mu \nu \alpha \beta} \,
T_{\alpha \beta}^e  \;.
\end{equation}

The second factor in Eq.~\eqref{matrix_element} is
\emph{universal} for all types of processes mediated by $s$ or
$t$-channel exchange of the massive  KK excitations. Let us
consider first the contribution to $\mathcal{S}$ from the
$s$-channel gravitons:
\begin{equation}\label{KK_sum}
\mathcal{S}(s) =  \frac{1}{\Lambda_{\pi}^2} \sum_{n=1}^{\infty}
\frac{1}{s - m_n^2 + i \, m_n \Gamma_n} \;.
\end{equation}
Here $\Gamma_n$ denotes the total width of the graviton with the
KK number $n$ and mass $m_n$~\cite{Kisselev:05_2}:
\begin{equation}\label{graviton_widths}
\frac{\Gamma_n}{m_n} = \eta \left( \frac{m_n}{\Lambda_{\pi}}
\right)^2 ,
\end{equation}
where $\eta \simeq 0.09$.

Note that the main contribution to sum~\eqref{KK_sum} comes from
the region $n \sim \sqrt{s}/\kappa \gg 1$. For such large values
of the KK number $n$, \emph{nonzero widths of the gravitons should
be taken into account}.

The sum in Eq.~\eqref{KK_sum} can be calculated analytically by the
use of the formula~\cite{Watson}
\begin{equation}\label{zeros_Bessel_sum}
\sum_{n=1}^{\infty} \frac{1}{ z_{n, \nu}^2 - z^2} = \frac{1}{2 z}
\, \frac{J_{\nu + 1}(z)}{J_{\nu}(z)} \; ,
\end{equation}
where $z_{n, \, \nu}$ ($n=1,2 \ldots$) are zeros of the function
$z^{-\nu} J_{\nu}(z)$. As a result, the following explicit
expression can be obtained from \eqref{KK_sum} (for details, see
Ref.~\cite{Kisselev:06}):
\begin{equation}\label{main_formula_final}
\mathcal{S}(s) = - \frac{1}{4 \bar{M}_5^3 \sqrt{s}} \; \frac{\sin 2A
+ i \sinh 2\varepsilon }{\cos^2 \! A + \sinh^2 \! \varepsilon } \; ,
\end{equation}
with the notations
\begin{equation}\label{parameter_A_epsilon}
A = \frac{\sqrt{s}}{\kappa} + \frac{\pi}{4}, \qquad \varepsilon  =
\frac{\eta}{2} \Big( \frac{\sqrt{s}}{\bar{M}_5} \Big)^3 \; .
\end{equation}

As one can see from \eqref{main_formula_final}, the magnitude of
$\mathcal{S}(s)$ is defined by $\bar{M}_5$ and $\sqrt{s}$,
\emph{not by} $\Lambda_{\pi}$, although the latter describes the
graviton coupling with the matter in the effective Lagrangian
\eqref{Lagrangian}.

The following inequalities result from \eqref{main_formula_final}:
\begin{equation}\label{bounds_ImS}
- \coth \varepsilon  \leqslant \mathrm{Im}
\,\mathcal{\tilde{S}}(s) \leqslant - \tanh \varepsilon  \;
\end{equation}
\begin{align}
\big| \mathrm{Re} \,\mathcal{\tilde{S}}(s) \big| &\leqslant
\frac{1}{1 + 2\sinh^2
\! \varepsilon } \; ,
\label{bounds_ReS}  \\
\left| \frac{\mathrm{Re} \, \mathcal{\tilde{S}}(s)}{\mathrm{Im}
\,\mathcal{\tilde{S}}(s)} \right| &\leqslant \frac{1}{\sinh 2
\varepsilon } \;,
\label{bound_ReS/ImS}
\end{align}
where the notation $ \mathcal{\tilde{S}}(s) = [2 \bar{M}_5^3
\sqrt{s} \,] \, \mathcal{S}(s)$ was introduced. Note that the
ratio $|\, \mathrm{Re} \, \mathcal{S}(s)/\mathrm{Im}
\,\mathcal{S}(s)|$ decreases with energy and becomes small at
$\sqrt{s} \simeq 3 \bar{M}_5$, while $\mathrm{Im}
\,\mathcal{S}(s)$ tends to  $(-1)/(2 \bar{M}_5^3 \sqrt{s})$.

Should one ignores the widths of the massive gravitons, and
replace a summation in KK number \eqref{KK_sum} by integration
over graviton masses,
\begin{equation}\label{sum_vs_int}
dn = \frac{\bar{M}_{\mathrm{Pl}}^2}{2 \pi \bar{M}_5^3} \; dm \;,
\end{equation}
he gets \cite{Kisselev:06,Giudice:04}:
\begin{equation}\label{KK_sum_zero_widths}
\mathrm{Im} \,\mathcal{S}(s) = - \frac{1}{2 \bar{M}_5^3 \sqrt{s}}
\;, \qquad \mathrm{Re} \, \mathcal{S}(s) = 0 \;,
\end{equation}
in contrast to the exact formula \eqref{main_formula_final}.

However, the series of low-massive resonances in the RS model with
the small curvature can be replace by the continuous spectrum
\emph{only in a trans-Planckian energy region}  $\sqrt{s} \gtrsim 3
\bar{M}_5$. It can be understood as follows. One may regard the set
of narrow graviton resonances to be the continuous mass spectrum
(within a relevant interval of $n$), if only
\begin{equation}\label{continuous_spectrum}
\Delta m_{\mathrm{KK}} < \Gamma_n
\end{equation}
is satisfied, where $\Delta m_{\mathrm{KK}}$ is the mass splitting.
As was shown in Ref.~\cite{Kisselev:06}, the inequality
\eqref{continuous_spectrum} is equivalent to the above mentioned
inequality
\begin{equation}\label{trans_planckian_region}
\sqrt{s} \gtrsim 3 \bar{M}_5 \; .
\end{equation}

We are working in another kinematical region, $\sqrt{s} \lesssim 3
\bar{M}_5$, since the 5-dimensional scale $\tilde{M}_5$ is assumed
to be equal to (or larger than) 1 TeV, while the collision energy is
fixed to be 1 TeV. That is why, for our calculations we will use the
analytical expression~\eqref{main_formula_final} which takes into
account both the discrete character of the graviton spectrum and
finite widths of the KK gravitons.

Our main formula~\eqref{main_formula_final} can be also applied to
the scattering of the brane particles, induced by exchanges of
$t$-channel gravitons~\cite{Kisselev:05,Kisselev:05_2}.  In the
kinematical region $\bar{M}_5^3/\kappa \gg - t \gg \kappa^2$, the
contribution from the $t$-channel gravitons looks
like~\cite{Kisselev:06}:
\begin{equation}\label{t_channel_final}
\mathcal{S}(t) = - \frac{1}{2\bar{M}_5^3 \sqrt{-t}} \; .
\end{equation}

Let us now apply our theoretical formulae
\eqref{main_formula_final}, \eqref{t_channel_final} for estimating
virtual graviton contributions to the process \eqref{process}. The
relations of cross sections with the quantities $\mathcal{S}(s)$
and $\mathcal{S}(t)$ can be found in Ref.~\cite{Giudice:04}.

Note first that at the LEP2 energy ($\sqrt{s} = 200$ GeV) the
gravity effects are very small with respect to the SM cross
section~\cite{Giudice:04,Kisselev:07}. Thus, the above mentioned
bound on $\bar{M}_5$ from LEP~\eqref{n=1_limit} should be also
applied to the 5-dimensional scale in our scheme.

In Fig.~\ref{fig:2.5TeV_1GeV} the graviton contribution is
presented as a function of the scattering angle of the final
electrons at the ILC energy $\sqrt{s} = 1$ TeV (solid line).
Another prediction is also shown (dashed line) which was
calculated under the assumption that the dense spectrum of the KK
gravitons can be approximated by a continuum~\cite{Kisselev:07}.

The next figure shows the dependence of the cross section ratios
on the gravity scale $\bar{M}_5$. Figs.~\ref{fig:1.8TeV_0.9GeV},
\ref{fig:1.8TeV_0.994GeV} demonstrate us that these ratios can be
very sensitive to small variations of the curvature $\kappa$
within a narrow region (0.9 GeV -- 1 GeV, in our case).

Let us note that the interference of gravity with the SM forces is
constructive in the Bhabha scattering, and the ratio
$\sigma(\mathrm{SM+grav})/\sigma(\mathrm{SM})$ is lager than 1.

Up to now, we considered the fixed collision energy. However, the
effects related with non-zero graviton widths remain significant
after energy smearing within some interval around average value of
$\sqrt{s}$, as one can see in Fig.~\ref{fig:bhabha_average}. Note
that the energy smearing has a small influence on the cross
section ratios in the kinematical region $\cos \theta <
-0.8$~\cite{Kisselev:07}.


\section{Conclusions}

In the present paper the contributions of the virtual $s$- and
$t$-channel KK gravitons to the Bhabha scattering at the ILC
energy $\sqrt{s} = 1$ TeV were estimated. We have considered the
small curvature option of the RS model with two branes ($\kappa
\ll \bar{M}_5$). In such a scheme, the KK graviton spectrum is a
series of the narrow low-mass resonances.

The 5-di\-mensional Planck scale $M_5$ was taken to be 1.8 TeV and
2.5 TeV, with the curvature parameter being restricted to the
small region $\kappa = 900 \mathrm{\ MeV} - 1 \mathrm{\ GeV}$,
that means $\sqrt{s} \sim M_5 \gg \kappa$. For numerical
estimates, we used formula~\eqref{main_formula_final}, which
describes the gravity contribution to the process-independent part
of the scattering amplitude $\mathcal{S}$~\eqref{matrix_element}.

\begin{figure}[htb]
\begin{center}
\resizebox{.6 \textwidth}{!}{\includegraphics{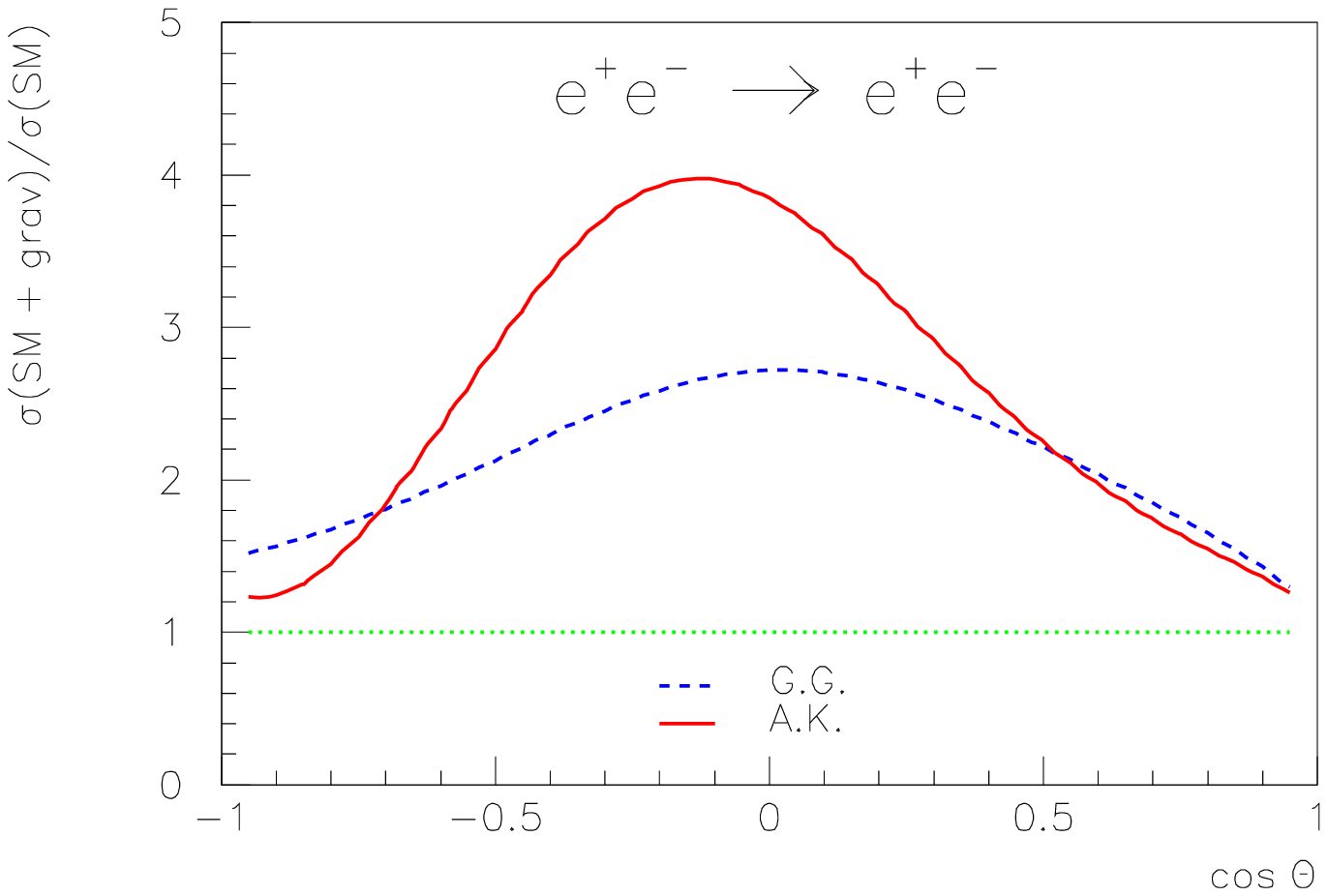}}
\caption{The correction to the Bhabha cross section resulting from
virtual graviton exchanges as a function of the scattering angle
at the ILC energy $\sqrt{s} = 1$ TeV.  The solid line is our
prediction which takes into account both a discrete character of
the spectrum and widths of the KK gravitons. The dashed line
corresponds to the continuous mass approximation for the graviton
spectrum. The parameters are $M_5 = 2.5$ TeV, $\kappa = 1$ GeV.}
\label{fig:2.5TeV_1GeV}
\end{center}
\end{figure}

\begin{figure}[htb]
\begin{center}
\resizebox{.6 \textwidth}{!}{\includegraphics{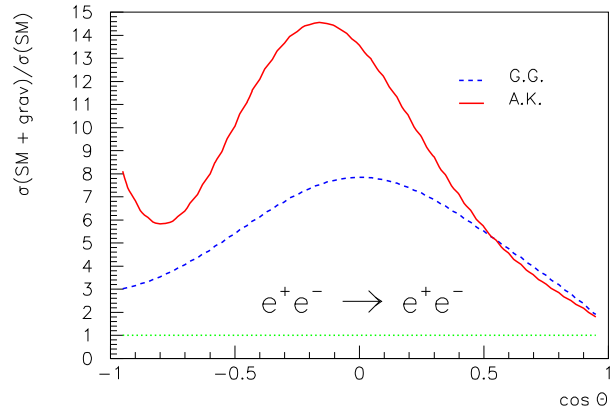}}
\caption{The same as in Fig.~\ref{fig:2.5TeV_1GeV}, with the
parameters $M_5 = 1.8$ TeV, $\kappa = 1$ GeV.}
\label{fig:1.8TeV_1GeV}
\end{center}
\end{figure}

\begin{figure}[htb]
\begin{center}
\resizebox{.6 \textwidth}{!}{\includegraphics{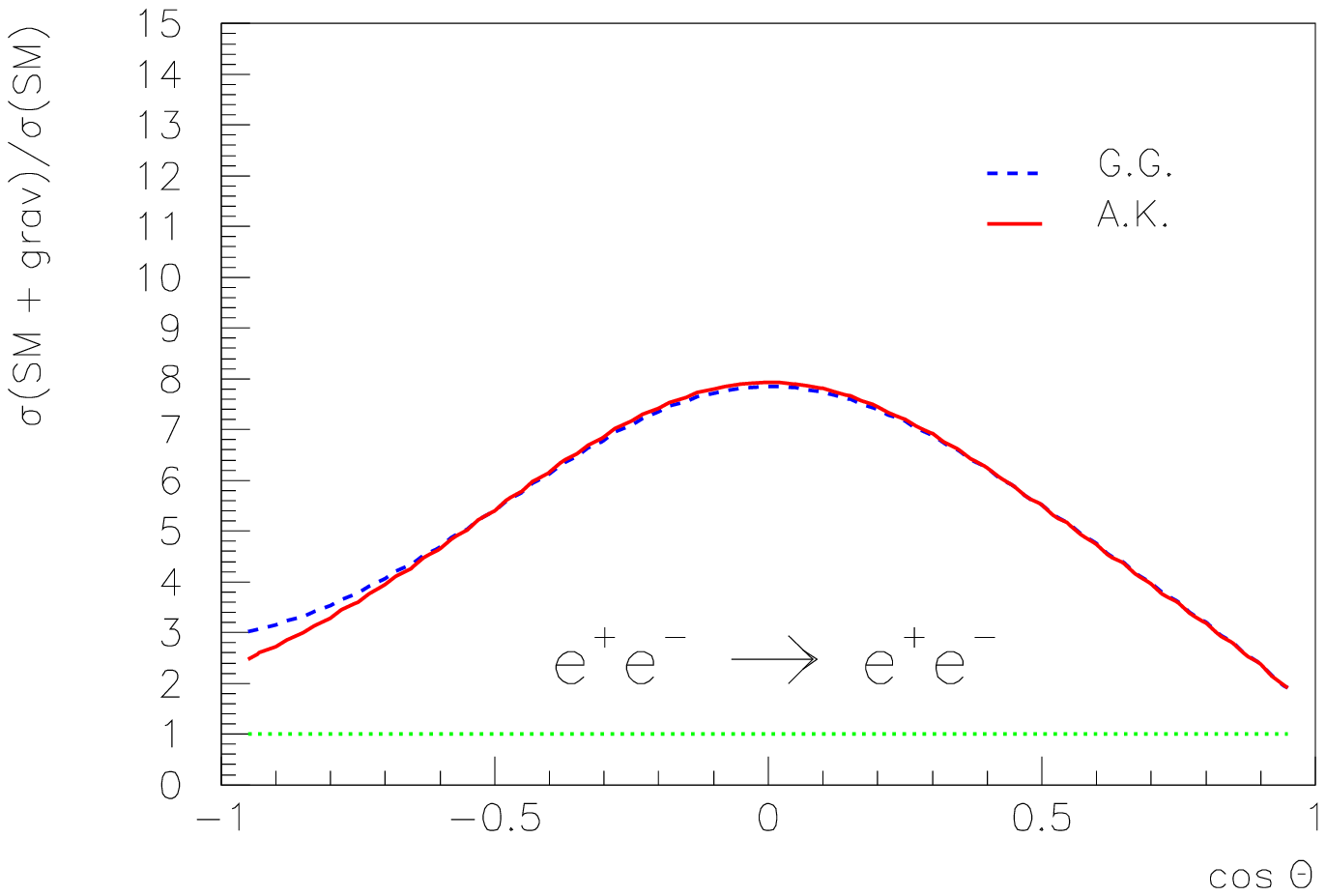}}
\caption{The same energy as in Fig.~\ref{fig:1.8TeV_1GeV}, except
for the parameter $\kappa = 900$ MeV.}
\label{fig:1.8TeV_0.9GeV}
\end{center}
\end{figure}

\begin{figure}[htb]
\begin{center}
\resizebox{.6 \textwidth}{!}{\includegraphics{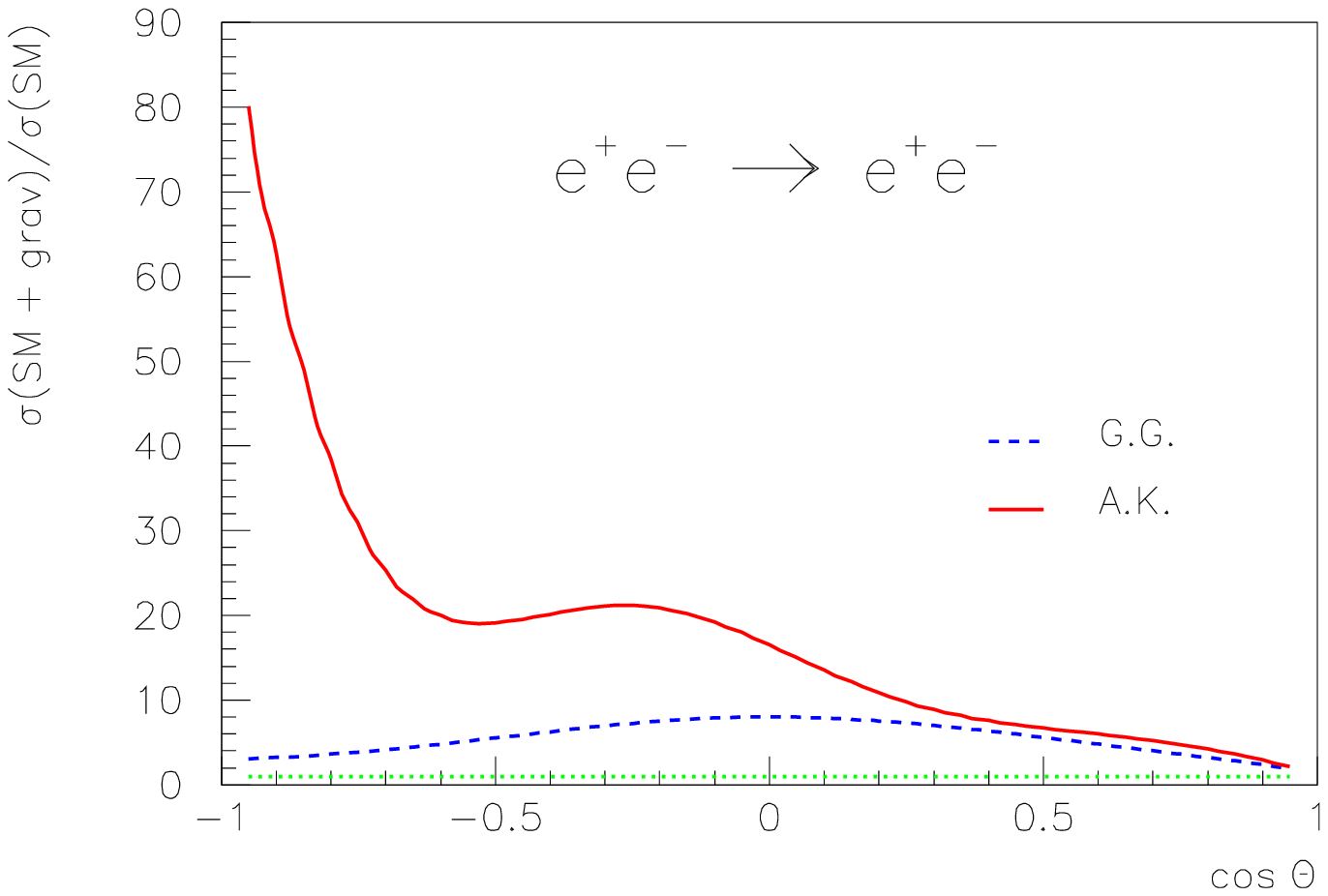}}
\caption{The same energy as in Fig.~\ref{fig:1.8TeV_1GeV}, except
for the parameter $\kappa = 994$ MeV.}
\label{fig:1.8TeV_0.994GeV}
\end{center}
\end{figure}

\begin{figure}[ht]
\begin{center}
\resizebox{.6\textwidth}{!}{\includegraphics{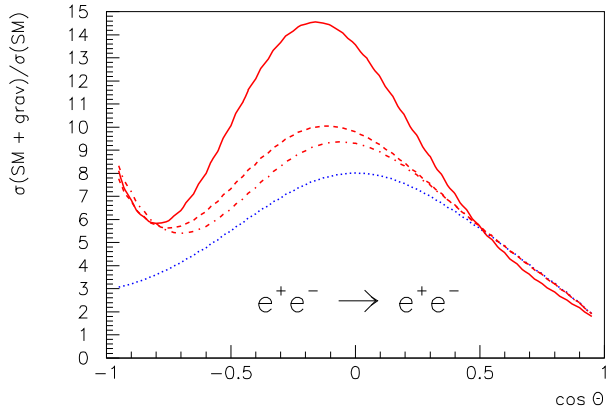}}
\caption{The dashed and dash-dotted lines correspond to smearing
of gravity cross section in the interval $(\sqrt{s} - \Delta
\sqrt{s}, \sqrt{s} + \Delta \sqrt{s})$, with $\Delta \sqrt{s} =
10$ GeV and $\Delta \sqrt{s} = 50$ GeV, respectively, while the
solid line corresponds to a case without energy smearing. The
average energy $\sqrt{s} = 1$ TeV, the parameters are the same as
in Fig.~\ref{fig:1.8TeV_1GeV}. The dotted line is obtained in zero
width approximation The number of resonances which lie within the
energy resolution is equal to 6 (31) for $\Delta \sqrt{s} = 10$
GeV ($50$ GeV).}
\label{fig:bhabha_average}
\end{center}
\end{figure}

For comparison, we have made calculations by using
Eqs.~\eqref{KK_sum_zero_widths} which treat the spectrum of the KK
graviton as a continuum. It was demonstrated that the sum in the
KK number can not be approximated by integration over graviton
mass. Moreover, the graviton widths should be properly taken into
account, as formula \eqref{main_formula_final} does. Our
predictions are presented in
Figs.~\ref{fig:2.5TeV_1GeV}-\ref{fig:bhabha_average}~\cite{Kisselev:07}.
Everywhere $\sigma(\mathrm{SM})$ means the SM differential cross
section with respect to $\cos \theta$, while
$\sigma(\mathrm{SM+grav})$ means the differential cross section
which takes into account both the SM and gravity interactions.

One should conclude that at fixed value of the fundamental gravity
scale $M_5$, the cross section ratio is very sensitive even to
slight variations of the curvature $\kappa$ (compare
Figs.~\ref{fig:1.8TeV_1GeV}, \ref{fig:1.8TeV_0.9GeV} and
\ref{fig:1.8TeV_0.994GeV}). The ratio
$\sigma(\mathrm{SM+grav})/\sigma(\mathrm{SM})$ can reach tens at
some values of the parameters (see
Figs.~\ref{fig:1.8TeV_0.994GeV}).

Such a behavior of the cross section ratios comes from the explicit
form of the function $\mathcal{S}(s)$~\eqref{main_formula_final}.
Indeed, at $\eta\,(\sqrt{s}/\bar{M}_5)^3 \ll 1$ the parameter
$\varepsilon$ in Eq.~\eqref{main_formula_final} is much less than 1,
and the function $\mathcal{S}(s)$ has a significant real part.%
\footnote{Remember that $\eta(\sqrt{s}/\bar{M}_5)^3 > 1$ in the
trans-Planckian region $\sqrt{s} > 3\bar{M}_5$.}
The interference with the SM contributions results in a nontrivial
dependence of $\sigma(\mathrm{SM+grav})/\sigma(\mathrm{SM})$ on
$\cos \theta$. Given the approximation with zero real part is used
\eqref{KK_sum_zero_widths}, no interference terms exist, and $\cos
\theta$-dependence of the cross section ratio becomes similar for
all sets of the parameters (dashed curves in
Figs.~\ref{fig:2.5TeV_1GeV}-\ref{fig:1.8TeV_0.994GeV}).

If the parameter $A = \sqrt{s}/\kappa + \pi/4$ in formula
\eqref{main_formula_final} obeys the equation $\cos A = 0$, the
denominator in \eqref{main_formula_final} becomes small, and,
correspondingly, the function $\mathcal{S}(s)$ becomes large. It
results in the rapid variation of the gravity contribution near
corresponding values of $\kappa$.

It is worth to note that both a discrete character of the mass
spectrum and nonzero widths of the KK gravitons are also important
in a model with a \emph{flat} compact extra
dimension~\cite{Kisselev:07}.


\section*{Acknowledgements}

I am grateful to G.F.~Giudice and V.A.~Petrov for fruitful
discussions.




\end{document}